\documentclass[a4paper,11pt]{article}

\usepackage{pos}
\usepackage{lipsum} 
\usepackage{nicefrac}
\usepackage{siunitx}
\usepackage{graphicx}
\usepackage{adjustbox}
\usepackage{caption}
\usepackage{lineno}

\title{Advanced Northern Tracks Selection using a Graph Convolutional Neural Network for the IceCube Neutrino Observatory}

\ShortTitle{Advanced Northern Tracks Selection for IceCube}

\author{The IceCube Collaboration \\{\normalsize \normalfont(a complete list of authors can be found at the end of the proceedings)}\\}

\emailAdd{soldin@physik.rwth-aachen.de}
\emailAdd{deng@physik.rwth-aachen.de}
\emailAdd{lasse.dueser@rwth-aachen.de}
\emailAdd{fuerst@physik.rwth-aachen.de}

\abstract{

The IceCube Neutrino Observatory is a cubic-kilometer detector located in the Antarctic ice at the geographic South Pole. It reads out over 5,000 photomultiplier tubes (PMTs) to detect Cherenkov light produced by secondary particles, enabling IceCube to identify both atmospheric and astrophysical neutrinos. One of the main challenges in this effort is effectively distinguishing between muons induced by neutrinos and those generated by cosmic-ray air showers. To address this challenge, the Advanced Northern Tracks Selection (ANTS) employs a graph convolutional neural network. This network is designed to utilize both the sensor data and the geometric arrangement of the detector's photomultiplier tubes (PMTs). By representing each module as a node in a graph and extracting features from each module, the network can capture and integrate both local and global features. This work details the implementation of the network architecture and highlights the improvements in background rejection efficiency compared to existing methods for selecting muon tracks.

\vspace{4mm}

{\bfseries Corresponding authors:}
Philipp Soldin$^{1*}$, 
Shuyang Deng$^{1}$, 
Lasse Düser$^{1}$, 
Philipp Fürst$^{1}$
\\
{$^{1}$ \itshape RWTH Aachen University}
\\[4mm]
$^*$ Presenter
}

\ConferenceLogo{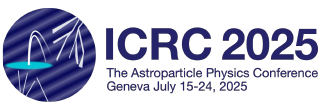}

\FullConference{39th International Cosmic Ray Conference (ICRC2025)\\
 15–24 July 2025\\
Geneva, Switzerland\\}

\begin{document}

\maketitle

\section{Introduction}
\noindent
The detection of high-energy astrophysical neutrinos by IceCube in 2013 \cite{astro_flux_discovery} marked a breakthrough in neutrino astronomy. Yet, key questions remain: What are the sources? What processes produce this flux, and what are its spectral characteristics?

\noindent
The IceCube Neutrino Observatory is well suited to answer such questions, instrumenting a cubic kilometer of Antarctic ice with \num{5160} digital optical modules (DOMs) on \num{86} strings, deployed at depths between \SI{1450}{\metre} and \SI{2450}{\metre}. These detect Cherenkov light from secondary particles resulting from neutrino interactions. For charged-current $\nu_{\mu}$ interactions, one of these secondaries is a muon, which creates a detectable, elongated light signal from Cherenkov emission emitted by the muon and its energy deposits along its trajectory -- a so-called \textit{track} signature in the detector.

\begin{figure}[b]
\centering
\includegraphics[width=0.9\linewidth]{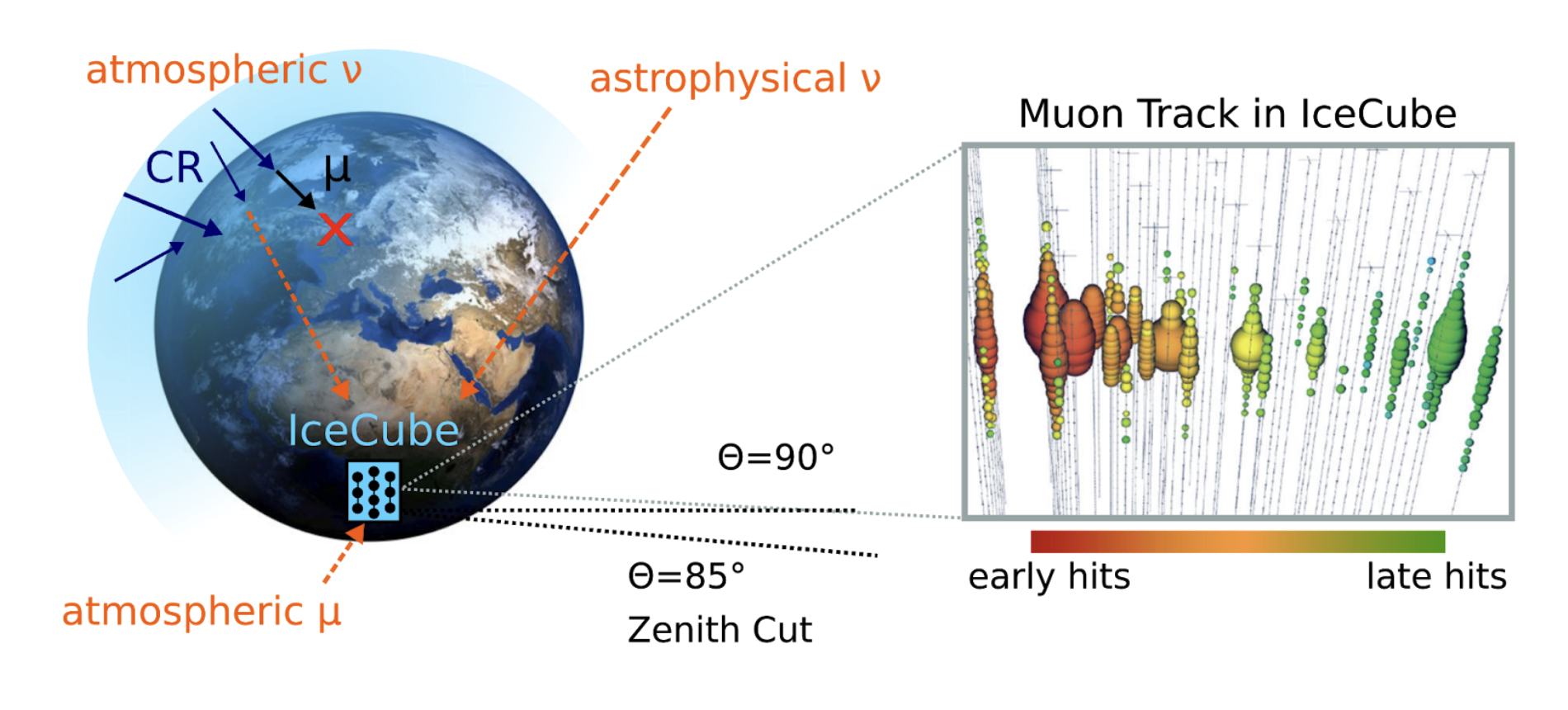}
\caption{Left: schematic of track-producing particle sources in IceCube. Atmospheric muons are the main background contribution for neutrinos and are removed using a cut at $\Theta > 85^\circ$ on the reconstructed zenith. Right: an exemplary muon track visualization; dot size indicates light intensity. The color indicates time, with red denoting early and green denoting late. Directional reconstruction achieves $<1^\circ$ median resolution above \SI{1}{\TeV}.}
\label{fig:intro_img}
\end{figure}

\noindent
Two main challenges arise when studying this kind of signal: First, low flux and cross-sections lead to inherently sparse statistics. This can be mitigated by including events with vertices outside the detector, utilizing the long range of muons. Second, a dominant background to neutrino-induced events originates from atmospheric muons resulting from cosmic ray showers, which appear as downgoing in the detector. This background contribution can already be greatly suppressed by selecting only upgoing events. Figure~\ref{fig:intro_img} shows a visualization of the described components and an example track event.

\noindent
The \textit{Northern Tracks} sample addresses both of these challenges by selecting well-reconstructed up-going track events, as only neutrinos can produce this signature in the detector. The main background contribution then arises from mis-reconstructed, air-shower-induced muons near the horizon. This sample provides high statistics and effective area, enabling results such as identifying NGC-1068 as a neutrino source~\cite{ngc_1068} and precise measurements~\cite{9.5yr_tracks} of the diffuse astrophysical neutrino flux spectrum. Over one million neutrino events have been collected over \num{13} years, with the majority of them being produced in the atmosphere. A high-statistics sample of atmospheric neutrinos enables an accurate description of this flux and has revealed effects such as seasonal flux variations \cite{seasonal_variations}.

\noindent
The selection begins with the IceCube muon track filter \cite{icecube_instrumentation_online_systems}, followed by reconstruction-based precuts. Initial steps select track-like events and already suppress the background from low-energy, down-going muons. Reconstructions with greater computational demands follow, which are used in a zenith cut $\Theta > 85^\circ$ for background rejection. Finally, two boosted decision trees (BDTs) are used to remove remaining backgrounds: one removes atmospheric muons that are close to the horizon or mis-reconstructed as up-going, while the other removes cascade events (spherical light emitted in muonless events). They use features related to reconstruction quality, event geometry, and event position in the detector. A final sample purity of over \SI{99.8}{\percent} is achieved~\cite{9.5yr_tracks}. 

\noindent
With more than a decade of data available, improving selection efficiency is a step toward increasing IceCube's efficiency, as higher statistics enhance the sensitivity of analysis. We demonstrate that replacing BDTs with a graph convolutional neural network (GCNN) using full DOM-level input instead of reconstructed high-level variables improves efficiency while maintaining purity. The new selection exhibits strong agreement between simulation and measurement on a small subsample of \SI{0.23}{years} of data collected in 2018.

\noindent
The training utilizes a comprehensive set of simulations, including various ice models, and encompasses approximately 20 million events in total. Neutrinos are modeled using NuGen (based on ANIS, \cite{nugen_anis}), while atmospheric muons are simulated with CORSIKA \cite{corsika}.

\noindent
We also present extensions of this architecture for various tasks, including a regression model for energy estimation and a classifier for event topology (e.g., throughgoing, starting, stopping, contained, skimming, and corner-clipping). The components of the presented architecture provide a flexible, general-purpose deep learning pipeline for IceCube track-like events.
\begin{figure}[h]
    \centering
    \includegraphics[width=0.6\linewidth, angle=-90]{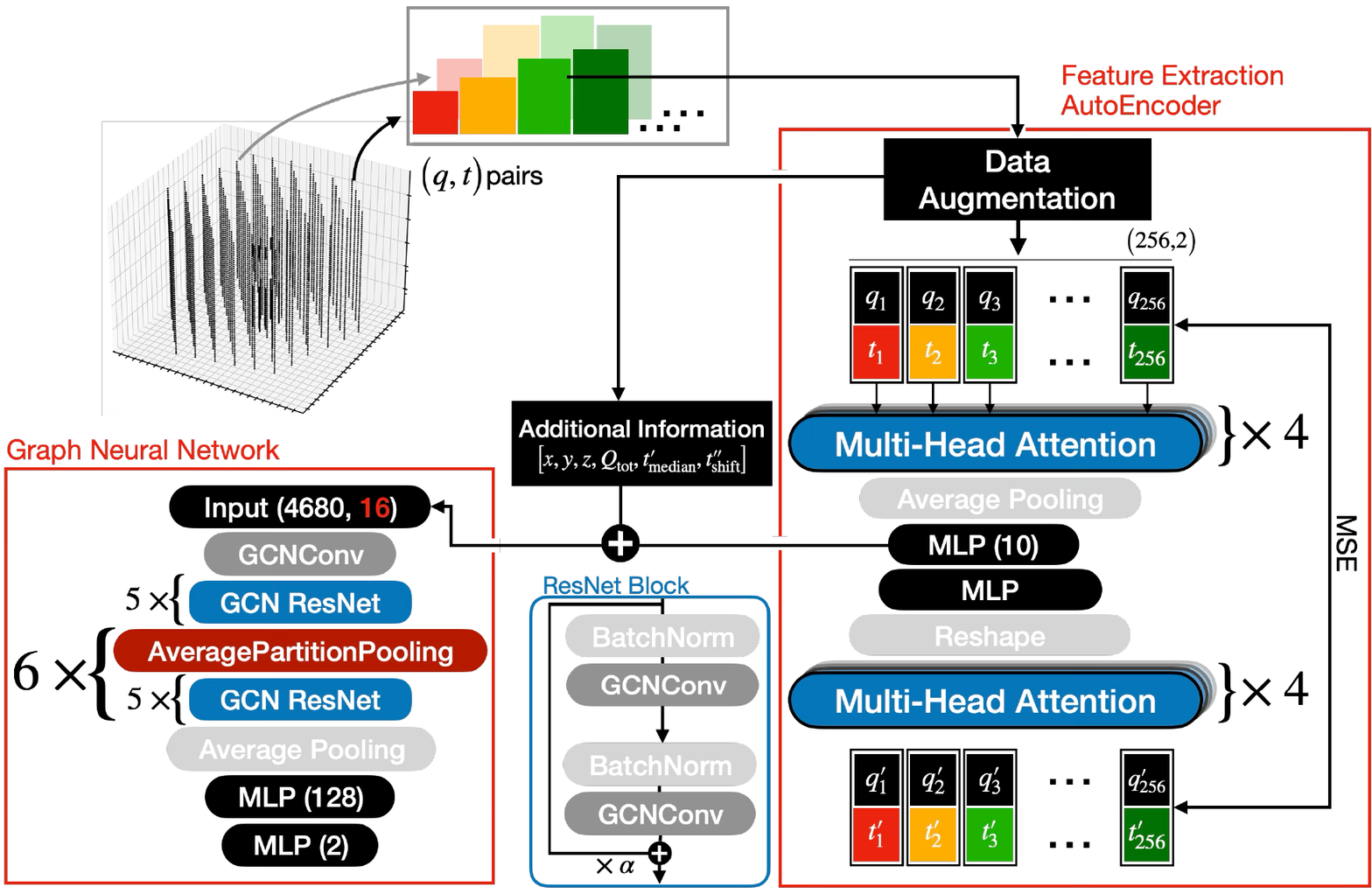}
    \caption{Depiction of the complete network architecture that can process full detector information using a transformer-based AutoEncoder and a Graph Convolutional Neural Network Architecture in a two-stage approach. The first $(q,t)$ pair detected by a DOM is represented in red and later $(q,t)$ pairs (from the same DOM) in green~\cite{vaswani2023attentionneed}. The Graph Neural Network then combines this information with additional information from each DOM in a Graph representing the entire detector.}
    \label{fig:network_ANTS}
\end{figure}

\section{Deep Learning Pipeline}
\noindent
The core of our event selection relies on a GCNN operating on features extracted from each DOM using a transformer-based encoder. This two-stage setup decouples feature extraction from event-level classification. The architecture is illustrated in Figure~\ref{fig:network_ANTS}.

\noindent
Each DOM records time series data of detected light pulses. Before encoding, reconstructed data undergoes an augmentation step. Reconstructed charge–time pairs within \SI{10}{\nano\second} are merged to reduce digitization and reconstruction artifacts that may differ between simulation and measurement. Features are then normalized by their respective standard deviations.

\noindent
The merged time values are further processed: the per-DOM median is subtracted to remove systematic offsets, followed by a signed square root transformation to standardize dynamic range and capture photon burst patterns. The series is then shifted so that the first hit occurs at time zero. The information about relative time information between DOMs is saved for later use in the network architecture.

\noindent
The transformer encoder is trained as part of an \textit{AutoEncoder} on this preprocessed data. The time and charge pairs for each DOM are mapped into an array that can hold up to \num{256} photon hits, sufficient for all IceCube event types. Masking ensures that the applied padding is ignored during training, which utilizes a mean squared error (MSE) loss function to compare the input and reconstructed sequences. The encoder consists of four multi-head attention layers~\cite{vaswani2023attentionneed} (each with \num{4} heads), global average pooling, and a multi-layer perceptron (MLP) that outputs \num{10} latent features. The decoder mirrors this with another MLP, a reshape operation, and four additional attention layers.

\begin{figure}
    \centering
    \includegraphics[width=0.48\linewidth]{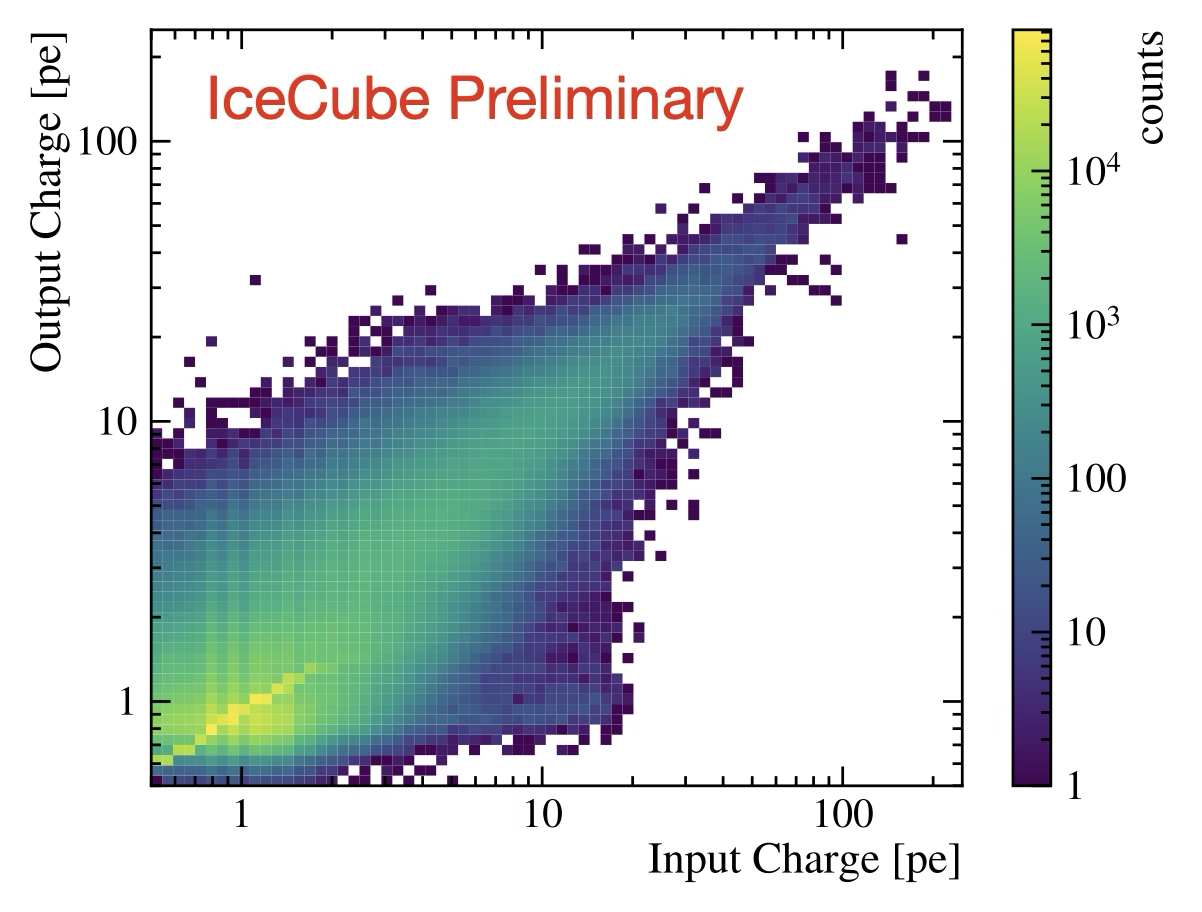}
    \includegraphics[width=0.48\linewidth]{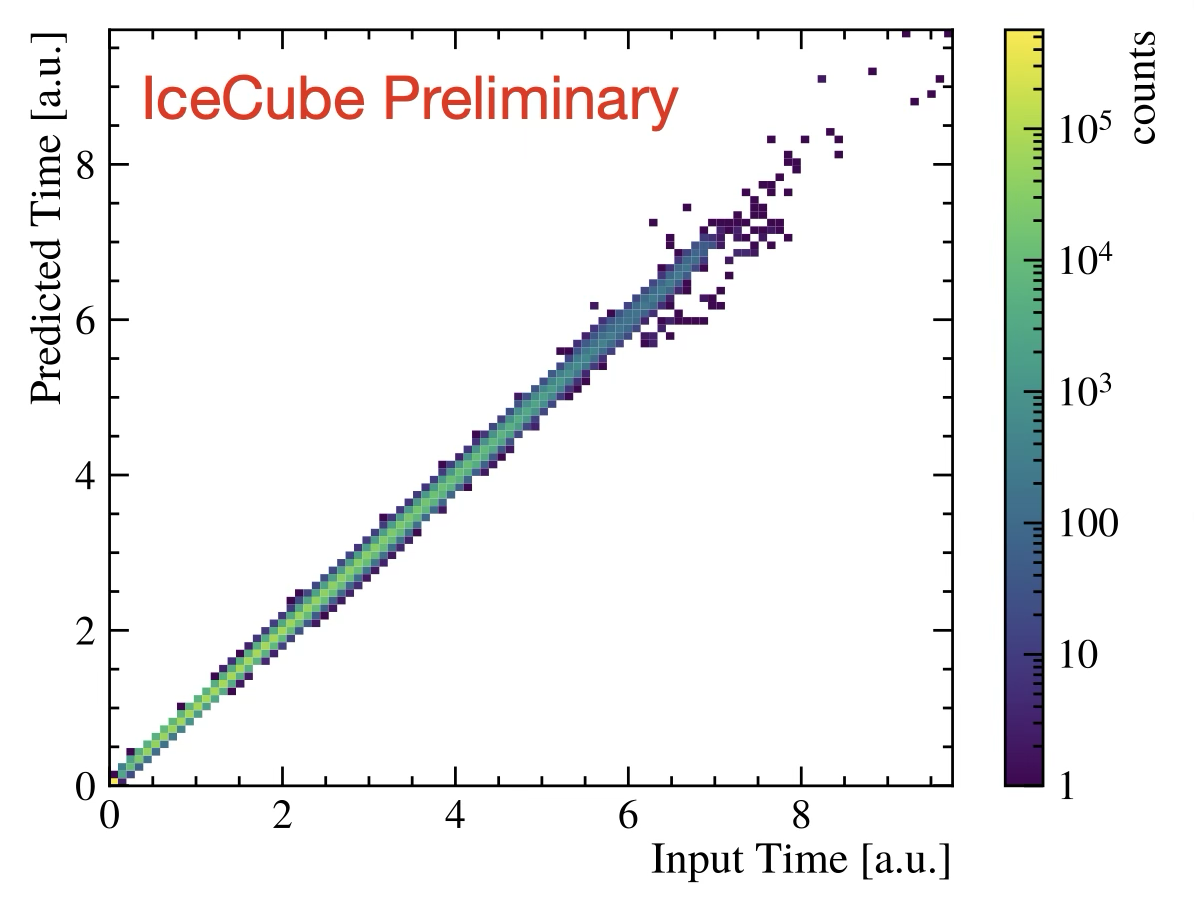}
    \includegraphics[width=0.48\linewidth]{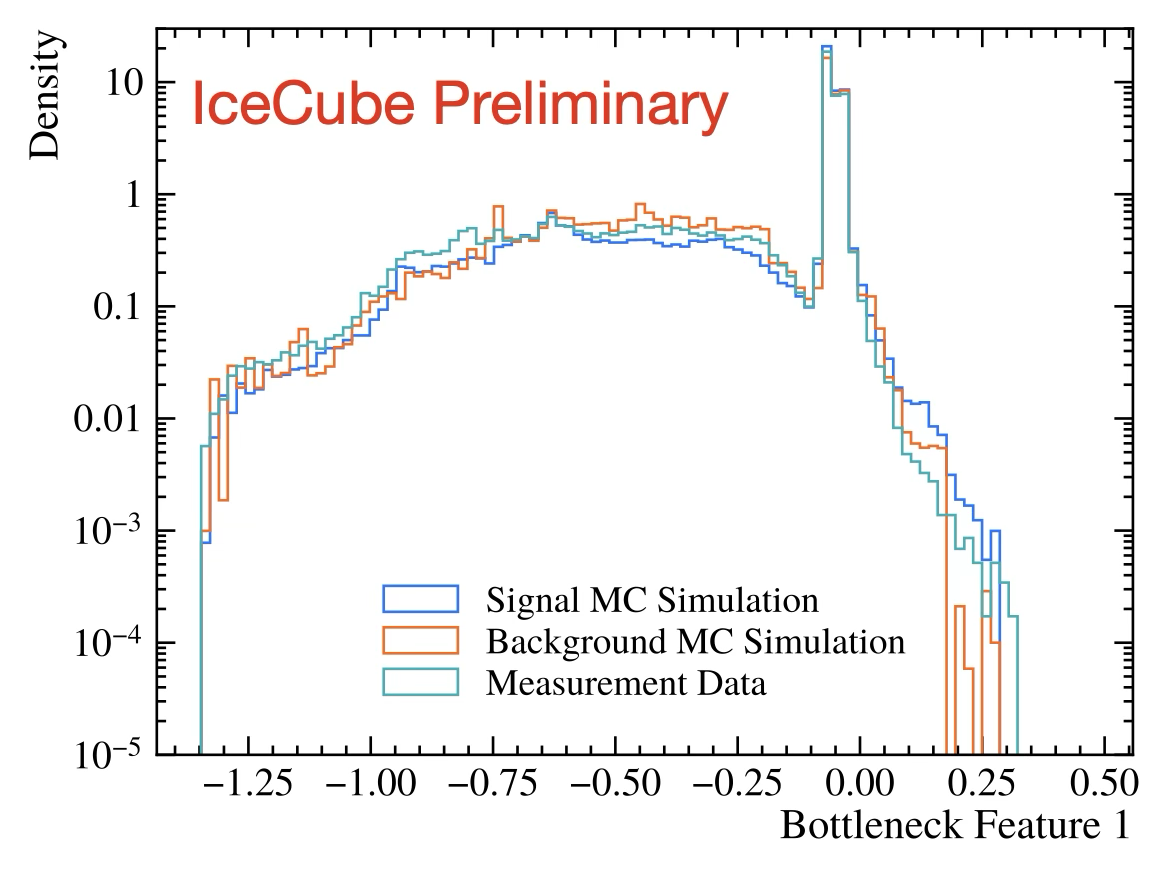}
    \caption{Comparison of the charge in photoelectrons (pe) and time in arbitrary transformed units (a.u.) with their respective reconstruction using a transformer-based AutoEncoder. Both input features show good agreement for signal and background simulation and measurement data, particularly in terms of time information. One exemplary AutoEncoder bottleneck feature is illustrated, showing good agreement between simulated and measured data, as well as in the AutoEncoder output.}
    \label{fig:input_data_AE}
\end{figure}

\noindent
As shown in Figure~\ref{fig:input_data_AE}, the AutoEncoder reconstructs timing with high accuracy for both simulated and measured data for the transformed times. Charges are also well reconstructed, though slightly less precise due to the lack of inherent ordering, and are given in photoelectrons (pe). The exemplary latent features in the third panel in Figure~\ref{fig:input_data_AE} show no differences across signal (Neutrino events simulated with NuGen), background (Muon events from CORSIKA Air shower simulations), or the sample measurement data, suggesting unbiased representations.

\noindent
Combined with the aforementioned auxiliary preprocessing features, such as relative time information between DOMs, the learned per-DOM representations serve as input to the GCNN for background rejection. The network (left part of Figure~\ref{fig:network_ANTS}) uses \num{16} per-DOM features and connects each DOM to its \num{16} nearest neighbors via a fixed, weighted adjacency matrix. GCNN layers are embedded in ResNet blocks~\cite{he2015deepresiduallearningimage} and followed by partition pooling~\cite{Bachlechner_2022}. The architecture is thus modeled on a typical structure of convolutional neural networks for images, whereas here, the pixels can assume arbitrary positions.

\noindent
After 70 GCNN layers in total, global average pooling and small MLPs were employed to produce the final classification on signal and background. The architecture is designed to handle complete event information and can be extended to additional tasks by replacing the final layers of the architecture. In addition to the background rejection task, we also demonstrate the ability to perform other tasks, such as topology classification or energy reconstruction, using the same base network. 

\section{Results}

\begin{figure}
    \centering
    \includegraphics[width=0.8\linewidth]{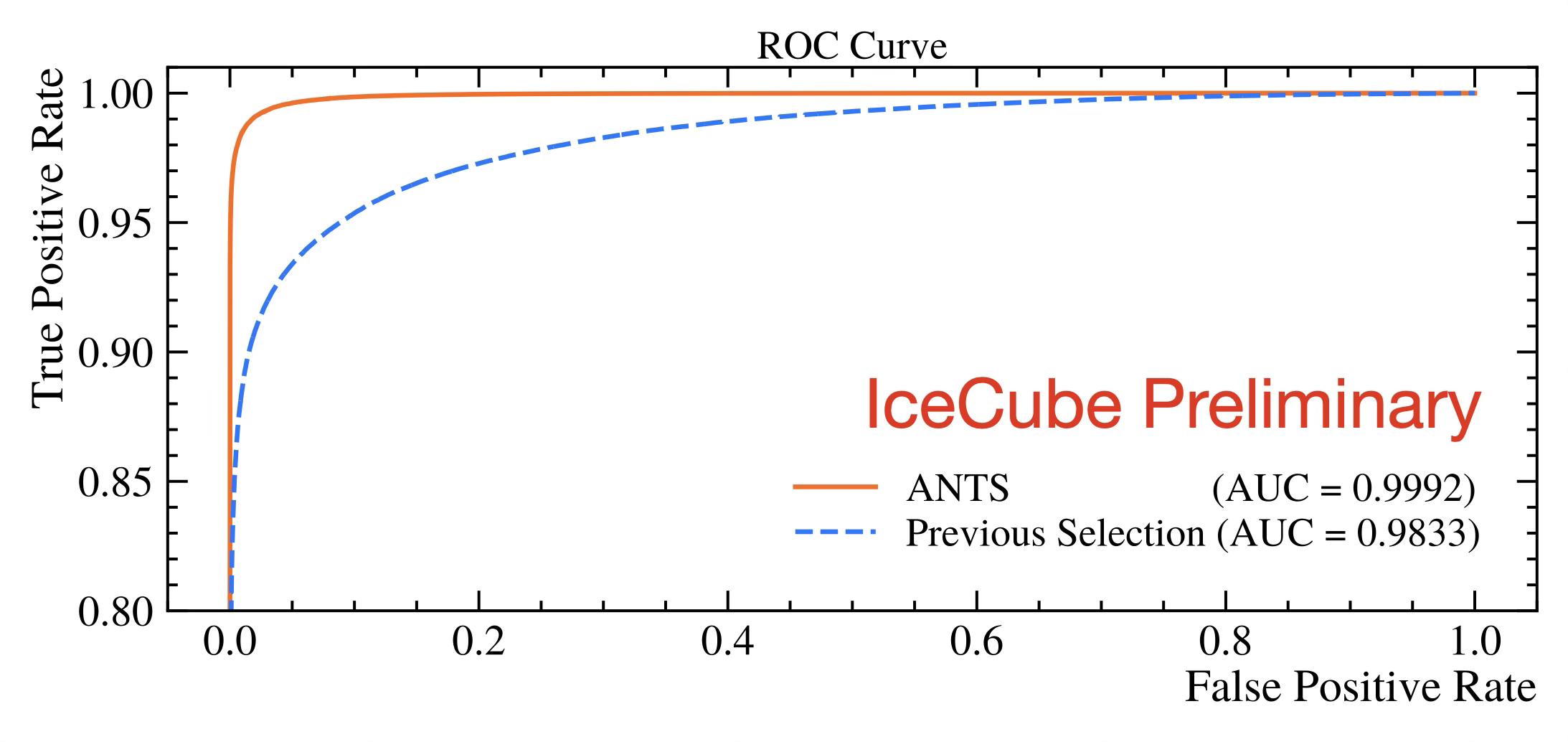}
    \caption{ROC Curve of the previous selection of muons from the northern hemisphere and the newly proposed ANTS selection, illustrating a drastic improvement in classification efficiency.}
    \label{fig:roc_curve}
\end{figure}

\noindent
The performance of the new event selection based on the graph neural network was evaluated in comparison to the previously used boosted decision tree (BDT) classifier. As shown in Figure~\ref{fig:roc_curve}, the new classifier significantly improves the separation between signal and background. The area under the ROC curve (AUC) increases from \num{0.9833} for the BDT to \num{0.9992} for the GCNN-based approach, demonstrating a significant improvement in background reduction.

\begin{figure}
    \centering
    \includegraphics[width=0.8\linewidth]{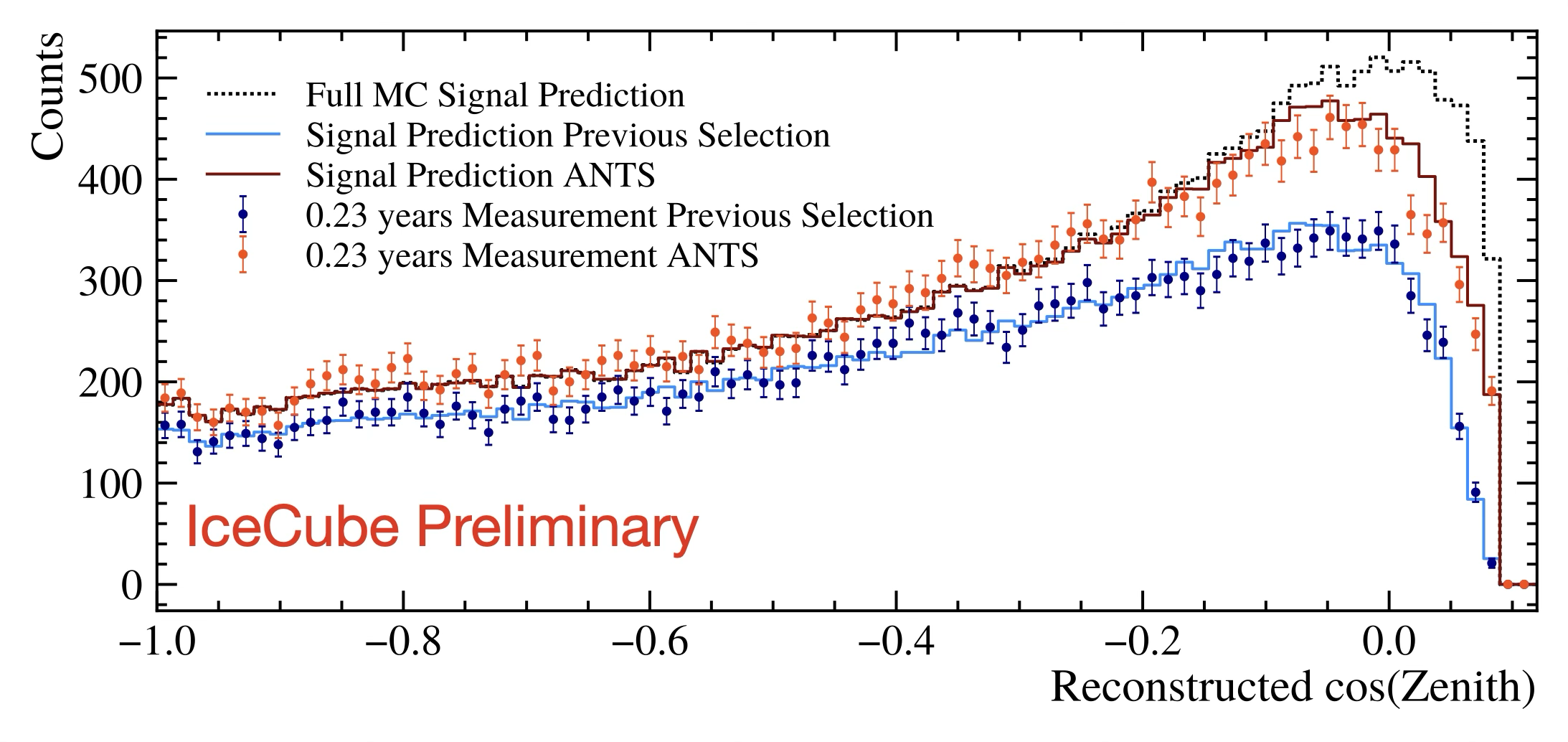}
    \caption{Improvement over the previous selection illustrated by the reconstructed $\cos\left(\mathrm{Zenith}\right)$ distribution. The ANTS selection recovers a total of $\sim 23000$ events in the \SI{0.23}{year} data sample with good Data/MC agreement while retaining \SI{95}{\percent} of the full signal prediction. This represents a $\sim\SI{25}{\percent}$ increase in event statistics compared to the BDT selection, which recovered $\sim 18500$ events from this sample.}
    \label{fig:zenith_ANTS_BDT_comparison}
\end{figure}

\noindent
A critical metric for physics analyses is the selection efficiency as a function of the reconstructed muon track direction. Figure~\ref{fig:zenith_ANTS_BDT_comparison} shows the distribution of the cosine of the reconstructed zenith angle after applying the selection cut chosen to match the original BDT purity of \SI{99.8}{\percent}. The new selection retains \SI{95}{\percent} of the original neutrino signal, indicated by the dotted line in Figure~\ref{fig:zenith_ANTS_BDT_comparison}, which enhances the selection by approximately \SI{25}{\percent}. This improvement in statistics is particularly valuable for subsequent analyses using the sample.
Effectively, the improved selection translates into gaining the equivalent of several additional years of data without requiring any new detector exposure. This is achieved primarily through the more effective utilization of the already recorded events.
The most challenging region for background rejection is near the horizon ($\cos\left(\mathrm{Zenith}\right)\approx 1$), where mis-reconstructed atmospheric muons are most likely to contaminate the sample. The new classifier enhances selection across the full zenith range, particularly in the upgoing region, where it achieves nearly \SI{100}{\percent} efficiency.
\noindent
The classification threshold was chosen to match the target purity of the previous selection. However, it is possible to further increase sample purity at the cost of signal efficiency, depending on the specific analysis requirements.
\noindent
Most importantly, the final selection shows excellent agreement between data and simulation across the relevant kinematic and reconstruction variables, as illustrated in Figure~\ref{fig:zenith_ANTS_BDT_comparison}. This agreement is critical, as the selected sample serves as the foundation for multiple downstream physics measurements and searches and was a particular focus during the development of the network and preprocessing pipeline.

\begin{figure}[ht]
    \centering
    \adjustbox{valign=m}{\includegraphics[width=0.5\textwidth]{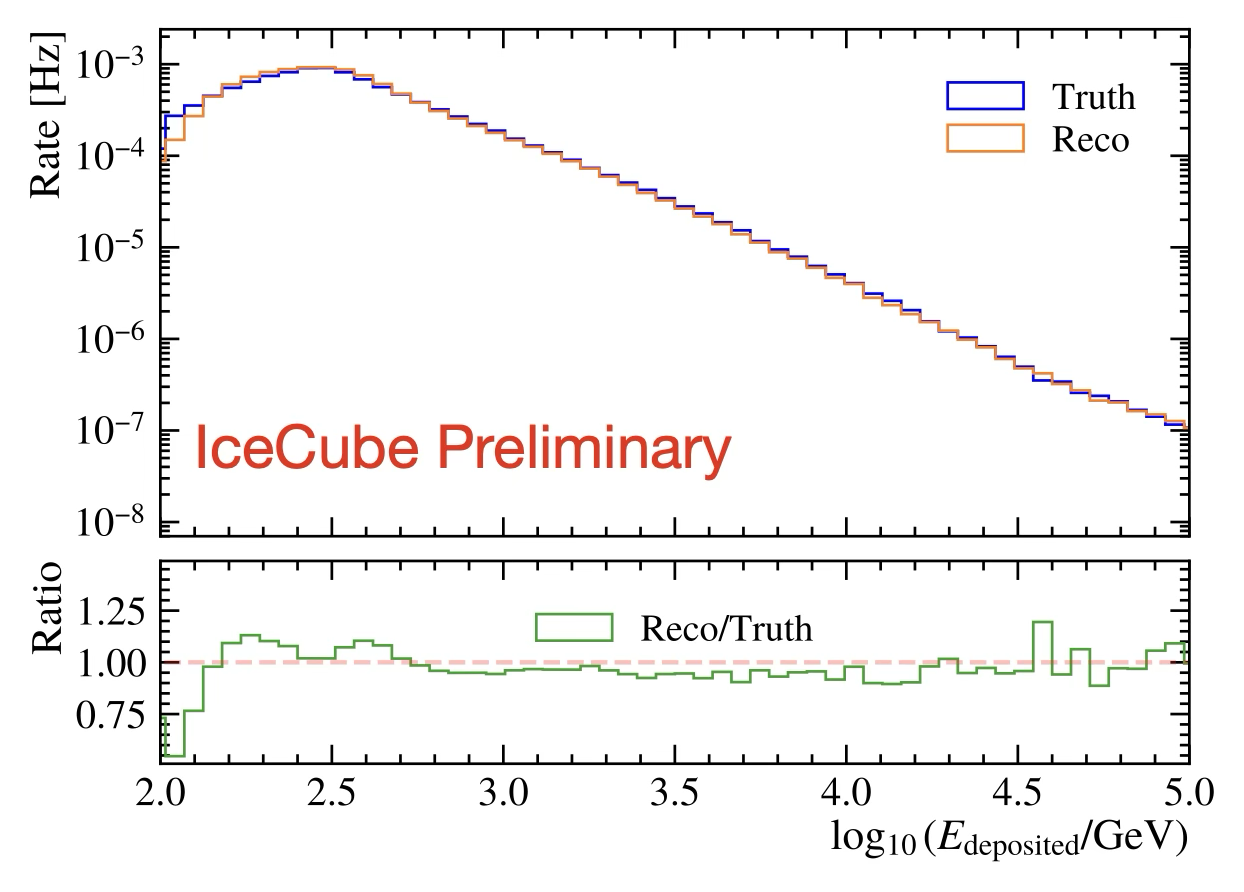}}%
    \hspace{0.05\textwidth}%
    \adjustbox{valign=m}{\includegraphics[width=0.45\textwidth]{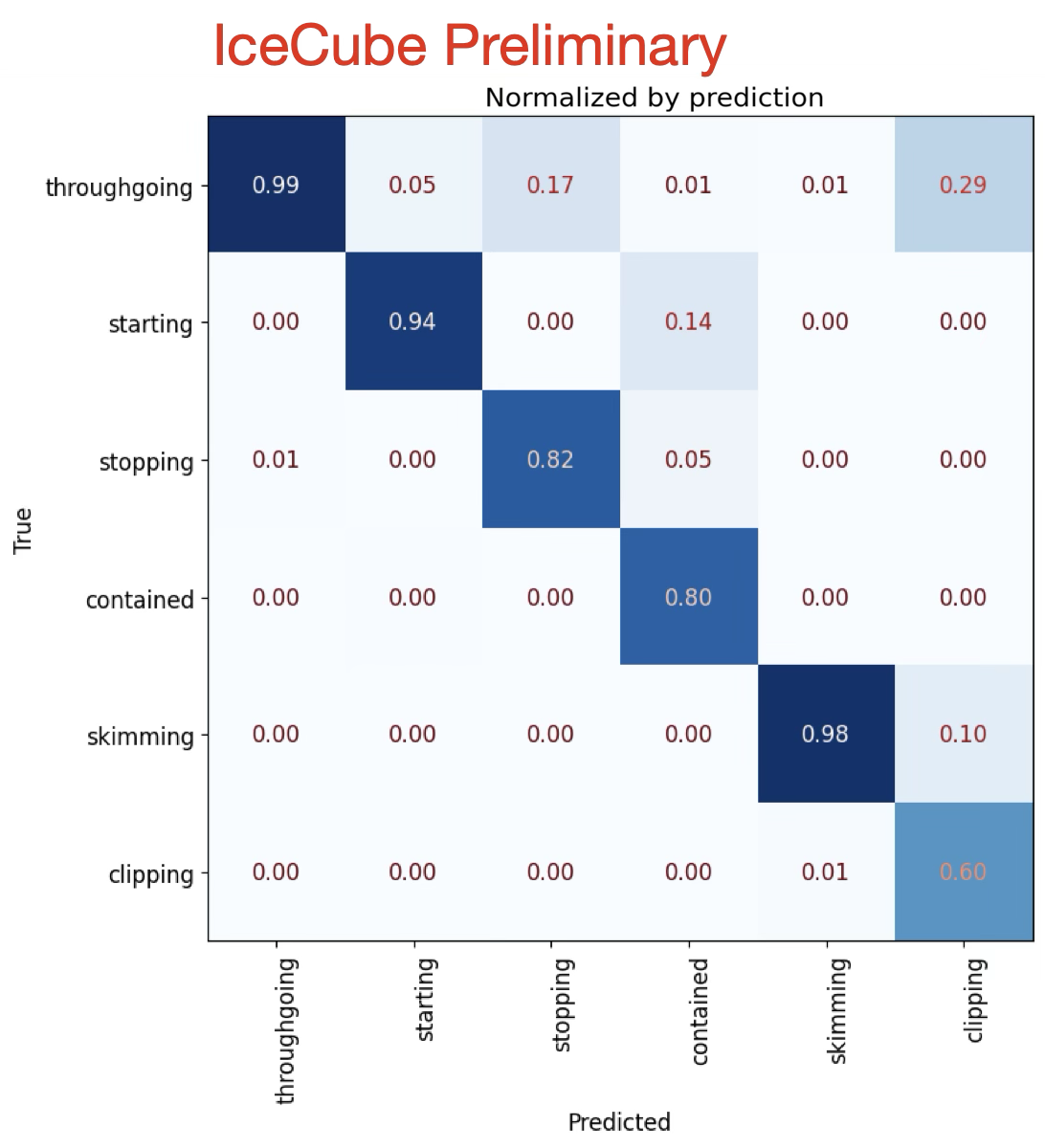}}%
    \caption{The left plot shows the result of the muon energy reconstruction, which demonstrates high agreement between the truth and reconstruction, as illustrated by the residual plot below. The right plot shows the classification confusion matrix between different event topologies, indicating that the ANTS architecture is capable of effectively distinguishing between event topologies.}
    \label{fig:aligned_images}
\end{figure}

\noindent
In addition to background reduction classification, the architecture can be trivially modified to support event-level regression and multi-class classification tasks. One such extension is muon energy reconstruction, where the network reconstructs the muons deposited energy in the detector. As shown in Figure~\ref{fig:aligned_images} (left), the regression model exhibits strong agreement between simulation and prediction. Minor deviations at low energies—likely due to limited statistics or resolution—are expected to diminish in future retraining.

\noindent
The second extension involves classifying the event topology, which describes how a muon event signature is detected within the detector. The network distinguishes between throughgoing, starting, stopping, contained, skimming, and corner-clipping events. The resulting confusion matrix, shown in Figure~\ref{fig:aligned_images} (right), demonstrates a good classification performance. Misclassifications are rare and mostly occur between classes that continuously merge into each other, consistent with the expected difficulty of these distinctions.

\noindent
These results highlight the network's flexibility as a general-purpose tool for various reconstruction and classification tasks within the IceCube experiment.


\section{Conclusion \& Outlook}
\noindent
We have presented a highly efficient classification network based on a graph neural network architecture that fully incorporates the spatial and temporal structure of IceCube events. By combining DOM-level features from a transformer-based encoder with graph-based reasoning, we developed a classifier for northern hemisphere muon tracks that achieves the same level of purity as the existing BDT-based selection while increasing the total selected statistics by approximately \SI{25}{\percent}.

\noindent
This gain in efficiency directly translates into enhanced sensitivity for a wide range of physics analyses. The network's ability to process complete event information makes it a powerful and flexible tool for other applications as well. We demonstrated its applicability beyond classification with tasks such as energy reconstruction and event topology classification.

\noindent
Looking forward, we plan to improve further the data–MC agreement using domain adaptation techniques based on backpropagation~\cite{ganin2015unsuperviseddomainadaptationbackpropagation}. These methods enforce consistency between simulation and real data by construction. Additionally, we aim to integrate the new selection into a complete physics analysis to validate the sample and quantify the physics reach. Systematic uncertainty studies will also be conducted to understand better the robustness of the network predictions and their impact on the resulting measurements. We also plan to evaluate the existing extensions and implement a neutrino directionality reconstruction to create a complete reconstruction package that can then be used for subsequent physics analyses.

\noindent
Given the network's fast evaluation time, we plan to explore applying it at earlier stages in the event selection chain, potentially even as part of real-time filtering. This would enable the earlier rejection of background events and improve the selection of rare signal events, thereby maximizing the utility of the IceCube data stream.
\bibliographystyle{ICRC}
\bibliography{references}

\clearpage

\section*{Full Author List: IceCube Collaboration}

\scriptsize
\noindent
R. Abbasi$^{16}$,
M. Ackermann$^{63}$,
J. Adams$^{17}$,
S. K. Agarwalla$^{39,\: {\rm a}}$,
J. A. Aguilar$^{10}$,
M. Ahlers$^{21}$,
J.M. Alameddine$^{22}$,
S. Ali$^{35}$,
N. M. Amin$^{43}$,
K. Andeen$^{41}$
% ...
\\
\\
% ...
$^{10}$ Universit{\'e} Libre de Bruxelles, Science Faculty CP230, B-1050 Brussels, Belgium \\
% ...
$^{16}$ Department of Physics, Loyola University Chicago, Chicago, IL 60660, USA \\
$^{17}$ Dept. of Physics and Astronomy, University of Canterbury, Private Bag 4800, Christchurch, New Zealand \\
% ...
$^{21}$ Niels Bohr Institute, University of Copenhagen, DK-2100 Copenhagen, Denmark \\
$^{22}$ Dept. of Physics, TU Dortmund University, D-44221 Dortmund, Germany \\
% ...
$^{35}$ Dept. of Physics and Astronomy, University of Kansas, Lawrence, KS 66045, USA \\
% ...
$^{39}$ Dept. of Physics and Wisconsin IceCube Particle Astrophysics Center, University of Wisconsin{\textemdash}Madison, Madison, WI 53706, USA \\
% ...
$^{41}$ Department of Physics, Marquette University, Milwaukee, WI 53201, USA \\
% ...
$^{43}$ Bartol Research Institute and Dept. of Physics and Astronomy, University of Delaware, Newark, DE 19716, USA \\
% ...
$^{63}$ Deutsches Elektronen-Synchrotron DESY, Platanenallee 6, D-15738 Zeuthen, Germany \\
$^{\rm a}$ also at Institute of Physics, Sachivalaya Marg, Sainik School Post, Bhubaneswar 751005, India \\
% ...

\subsection*{Acknowledgments}

\noindent
The authors gratefully acknowledge the support from the following agencies and institutions:
USA {\textendash} U.S. National Science Foundation-Office of Polar Programs,
U.S. National Science Foundation-Physics Division,
U.S. National Science Foundation-EPSCoR,
U.S. National Science Foundation-Office of Advanced Cyberinfrastructure,
Wisconsin Alumni Research Foundation,
Center for High Throughput Computing (CHTC) at the University of Wisconsin{\textendash}Madison,
Open Science Grid (OSG),
Partnership to Advance Throughput Computing (PATh),
Advanced Cyberinfrastructure Coordination Ecosystem: Services {\&} Support (ACCESS),
Frontera and Ranch computing project at the Texas Advanced Computing Center,
U.S. Department of Energy-National Energy Research Scientific Computing Center,
Particle astrophysics research computing center at the University of Maryland,
Institute for Cyber-Enabled Research at Michigan State University,
Astroparticle physics computational facility at Marquette University,
NVIDIA Corporation,
and Google Cloud Platform;
Belgium {\textendash} Funds for Scientific Research (FRS-FNRS and FWO),
FWO Odysseus and Big Science programmes,
and Belgian Federal Science Policy Office (Belspo);
Germany {\textendash} Bundesministerium f{\"u}r Forschung, Technologie und Raumfahrt (BMFTR),
Deutsche Forschungsgemeinschaft (DFG),
Helmholtz Alliance for Astroparticle Physics (HAP),
Initiative and Networking Fund of the Helmholtz Association,
Deutsches Elektronen Synchrotron (DESY),
and High Performance Computing cluster of the RWTH Aachen;
Sweden {\textendash} Swedish Research Council,
Swedish Polar Research Secretariat,
Swedish National Infrastructure for Computing (SNIC),
and Knut and Alice Wallenberg Foundation;
European Union {\textendash} EGI Advanced Computing for research;
Australia {\textendash} Australian Research Council;
Canada {\textendash} Natural Sciences and Engineering Research Council of Canada,
Calcul Qu{\'e}bec, Compute Ontario, Canada Foundation for Innovation, WestGrid, and Digital Research Alliance of Canada;
Denmark {\textendash} Villum Fonden, Carlsberg Foundation, and European Commission;
New Zealand {\textendash} Marsden Fund;
Japan {\textendash} Japan Society for Promotion of Science (JSPS)
and Institute for Global Prominent Research (IGPR) of Chiba University;
Korea {\textendash} National Research Foundation of Korea (NRF);
Switzerland {\textendash} Swiss National Science Foundation (SNSF).

\end{document}